\documentclass[fleqn,12pt,twoside]{article}
\usepackage{espcrc1}
\usepackage{graphicx}
\newcommand \Pomeron {I\!\!P}
\title{Coherence Effects in  Diffractive
Electroproduction from Nuclei}

\author{L. Frankfurt,\address{School of Physics and Astronomy,
Raymond and Beverly Sackler,\\ Faculty of Exact Science, 
Tel Aviv University, Ramat Aviv 69978, Tel Aviv , Israel}
         M. Strikman,\address{ Pennsylvania State University,
 University Park, Pennsylvania 16802, USA}
         M. Zhalov\address{Petersburg Nuclear Physics Institute,
 Gatchina 188350, Russia}}
\begin{document}
\maketitle

\begin{abstract}
We argue that study of the cross section of coherent photo(electro)
production of $\rho,\rho^{\prime}, ...$-mesons provides an effective 
method to probe onset of black body limit(BBL) in the soft and hard QCD
interactions. We illustrate the expected features of the onset of BBL
using generalized vector dominance model(GVDM). We show that this model
describes very well $\rho$-meson coherent photoproduction at $6 \leq
E_{\gamma} \leq 10$ GeV. The advantages of the process of coherent dijet
production and hard diffractive processes in general for probing the  
onset of  BBL and measuring the light-cone wave function of the photon
in a hard scattering regime where decomposition over twists becomes 
inapplicable are explained. We argue also  that the regime of color 
fluctuations manifests itself at intermediate energies in exclusive and 
semiexclusive photoproduction of vector mesons and that the color 
transparency(CT) should reveal itself in these processes at $Q^2$ of the 
order of a few GeV$^2$.
\end{abstract}

\section{INTRODUCTION}
Studies of the coherent interactions of photons with nucleons and
nuclei were one of the highlights of the strong interaction studies of
the seventies, for a review see \cite{yenn}. New fundamental  questions to 
be investigated in the coherent processes are how interactions depend on the 
type of the projectile with increase of the size/thickness of the target and
how to measure various components of the light cone wave function of the 
photon. Several regimes appear possible. In the case of a hadronic projectile
(proton, pion, etc)  high-energy interactions with the nucleus rather rapidly 
approach a black body limit in which the total cross section of the 
interaction is equal to $2\pi R_A^2$. Another extreme limit is the interaction
of small size projectiles (or wave packages). In this case in a wide range of 
high energies the system remains frozen during the passage through  the 
nucleus and the regime of color transparency is reached in which the 
interaction of the small size projectile with a nucleus is rather weak.  
The amplitude of interactions is  proportional to the product of the matrix 
element of the two dimensional Laplace operator  between the light-cone  wave 
functions of projectile and hadronic final state and the gluon density of the 
nucleus which is somewhat smaller than the sum of the nucleon gluon  densities
due to the leading twist nuclear shadowing. 
However the cross section rapidly  grows 
with energy reflecting the fast increase of the gluon densities at small
${\it x}$ and large $Q^2$.
Then the interaction may ultimately  reach the black limit 
from the perturbative domain corresponding to quite
different pQCD dynamics, in particular, it could be reached 
already at ${\it x}\geq 10^{-3}$ where  $\ln {\it x} $ effects are a small
correction. The BBL for the interaction of the small size dipoles with heavy 
nuclei represents a new regime of interactions  when the leading twist 
approximation and therefore the whole notion of the parton distributions 
becomes inapplicable for the description of hard QCD processes in the small 
${\it x}$ regime.
Obviously there should also exist  many cases when the projectile represents 
a superposition of configurations of different sizes that leads to 
fluctuations of the interaction strength. In this respect interactions of real
and virtual photons with heavy nuclei provide unique opportunities since the 
photon wave function contains both the hadron-like configurations (vector 
meson dominance) and the direct photon configurations (small $q\bar q$ 
components). The important advantage of the  photon is that at high energies 
the BBL is manifested in diffraction into a multitude of the hadronic final 
states  (elastic diffraction $\gamma \to \gamma $ is negligible)
while in the hadron case only  elastic diffraction survives  in the BBL
and details of the dynamics leading to this regime remain hidden.
Spectacular manifestations of BBL in (virtual) photon diffraction include
strong enhancement of the large mass tail of the diffractive spectrum as
compared to the expectations of the triple Pomeron limit,  large cross
section of the high $p_t$ dijet production \cite{BBL}.
Investigation of the coherent diffraction in BBL would allow to perform
unique measurementws of various components of the light cone wave function of 
the photon, providing a much more detailed information than similar 
measurements in the  regime where leading twist dominates.

In preQCD time V.Gribov \cite{Gribov} explored the  complete absorption of 
hadrons by heavy nucleus to calculate the  total cross section of
photo(electro)production processes off heavy nuclei through the hadron
polarization operator for the photon. The distinctive feature of Gribov's 
approach is that the contribution of large masses in the wave function of 
projectile photon (a direct photon contribution)  is not suppressed. 
Consequently, the photon with energy  $q_0=\omega_{\gamma}$ interacts with 
a nucleus with the total cross section  
$\sigma_{\gamma A}^{tot}\propto  2\pi R_A^2\alpha_{em} 
\ln (2q_0/R_Am_{\rho}^2)$ 
for $A\gg 1$. 
This expression
grossly violates expectations of the Bjorken scaling for the $Q^2$ 
dependence of $\sigma_{\gamma A}^{tot}$ and is qualitatively different from 
the hadron case where $\sigma_{hA}^{tot} \approx  2\pi R_A^2$. 
To overcome this puzzle  J.Bjorken suggested a long time ago the aligned jet 
model in which only $q\bar q$ pairs with small $p_t$  can interact while high 
$p_t $ configurations in the photon wave function remain sterile\cite{bj}. 
Existence of sterile states has been explained later as due to the color 
transparency phenomenon \cite{FS88}.
More recently it was  understood that  some states which behave as  sterile 
at moderate energies, interact at high enough energies with a hadron
target with  cross sections comparable to that for soft QCD phenomena
\cite{BBFHS}.
Thus the Gribov's assumptions are justified in QCD for the interaction
of a range of  hadronic components of the photon wave function with heavy
nucleus target.  At the same time even at small ${\it x}$ the color 
transparency still survives for some  components and one needs smaller 
${\it x}$ to reach the BBL than that studied so far experimentally in 
$ep$ collisions. It is worth emphasizing that the hypothesis of BBL 
corresponds to the assumption that at sufficiently small ${\it x}$ partons 
with large virtuality interact with heavy nuclei without any suppression 
with a cross sections $\approx  2\pi R_A^2$. Namely this feature of the BBL 
is responsible for the gross violation of the Bjorken scaling and for 
the above mentioned qualitative difference of the energy dependence of 
$\sigma_{tot}^{\gamma A} $ and  $\sigma_{tot}^{h A}$.

In this talk we  summarize our recent studies of the various regimes of
the coherent photo/electro production off nuclei \cite{BBL,FSZpsi,FSZrho}:
the onset of the BBL regime, phenomenon of color transparency and perturbative
color opacity related to the leading twist nuclear gluon shadowing, and the 
pattern of soft QCD phenomena in the proximity to the black body limit.
We find that the  onset of color fluctuations occurs in the photoproduction,
while the onset of the color transparency in the exclusive electroproduction
is expected already at intermediate energies with increase of $Q^2$. 
\section{VECTOR MESON PRODUCTION OFF NUCLEI IN
 THE GENERALIZED VECTOR DOMINANCE MODEL}
We have used  generalized vector dominance model(GVDM)
\cite{Gribov,Brodsky} to describe coherent photoproduction of hadronic
states of $M\leq 2$ GeV off nuclei and consider the onset of BBL in
the soft regime. The main ambiguity in generalizing  the vector dominance 
model is the  issue of nondiagonal transitions where a photon initially 
converts into one vector state - $V_1$ which through diffractive interactions 
with a nucleon converts into another state $V_2$. Such amplitude  interferes 
with the process of direct production of $V_2$. The importance of the
nondiagonal transitions could be justified on the basis of the interpretation 
of the early Bjorken scaling for moderately small ${\it x}\sim 10^{-2}$
as due to the color transparency phenomenon - presence in the virtual photon 
of hadron type and point-like type configurations  \cite{FS88}.  These 
amplitudes are  also crucial for ensuring a quantitative matching with 
perturbative QCD  regime for $Q^2\leq$ ~few GeV$^2$\cite{FGS97}.
Hence it is reasonable to use GVDM  for  the modeling of the
production of the light states off nuclei.

The amplitude of the vector meson production off a nucleon
can be written within the GVDM  as
\begin{equation}
A(\gamma + N\to V_j + N )=\sum_{i} {e\over f_{V_i}}A(V_i + N \to V_j + N),
\label{GVDM}
\end{equation}
where $f_{V_i}$ are expressed through 
the width of the $V_{i}\to e^+e^-$ decay. In the case of nuclei
calculation of the vector meson production amplitude within the 
Glauber approximation requires taking into  account both  the nondiagonal 
transitions due to the transition of the photon to a different meson
$V'$ in the vertex $\gamma \to V' $ and due to change of the meson in
multiple rescatterings like $\gamma \to V \to V' \to V$. This physics
is equivalent to inelastic shadowing phenomenon familiar from
hadron-nucleus scattering \cite{Gribovinel}.
The Glauber model for the description of these processes is well known, see
e.g. \cite{yenn} \footnote{In our  calculation we neglect the triple
Pomeron contribution which is present at high energies. This
contribution though noticeable for the scattering off the lightest nuclei
becomes a very small correction for the scattering of heavy nuclei due
the strongly absorptive nature of interaction at the central impact 
parameters.}.
In Ref. \cite{Pautz:qm} the simplest nondiagonal model
was considered with two states $\rho$ and $\rho^{\prime}$ which
have the same diagonal amplitudes of scattering off a nucleon
and the fixed ratio of coupling constants:
$f_{\rho^{\prime}}/f_{\rho}=\sqrt{3}$,
while the ratio of the nondiagonal and diagonal amplitudes
$A(\rho + N \to \rho^{\prime} +N)/A(\rho + N \to \rho +N)=-\epsilon$,
and the value $\sigma^{tot}_{\rho N}$ were found from the fit to
the forward $\gamma +A\to \rho +A$ cross sections
measured at $\omega_{\gamma}=$6.1, 6.6 and 8.8 GeV\cite{MIT}.
This model with reasonable values of $\sigma_{\rho N}$ and $\epsilon$  
allowed to bring the value of $f_{\rho}$ determined from the photoproduction 
of $\rho$-mesons off protons at assumption of approximate equality of the 
cross sections of $\rho-N$ and $\pi -N $ interactions into a good agreement 
with the value extracted from the $e^+e^-$ data thus removing a long standing
20\% discrepancy between two determinations. One should emphasize here that 
in such VDM extension $\rho^{\prime}$-meson approximates the hadron production
in the interval of hadron masses $\Delta M^2 \sim 2\, GeV^2$, so the values 
of the production cross section refer to the corresponding mass interval.

\begin{figure}
\includegraphics[scale=0.8]{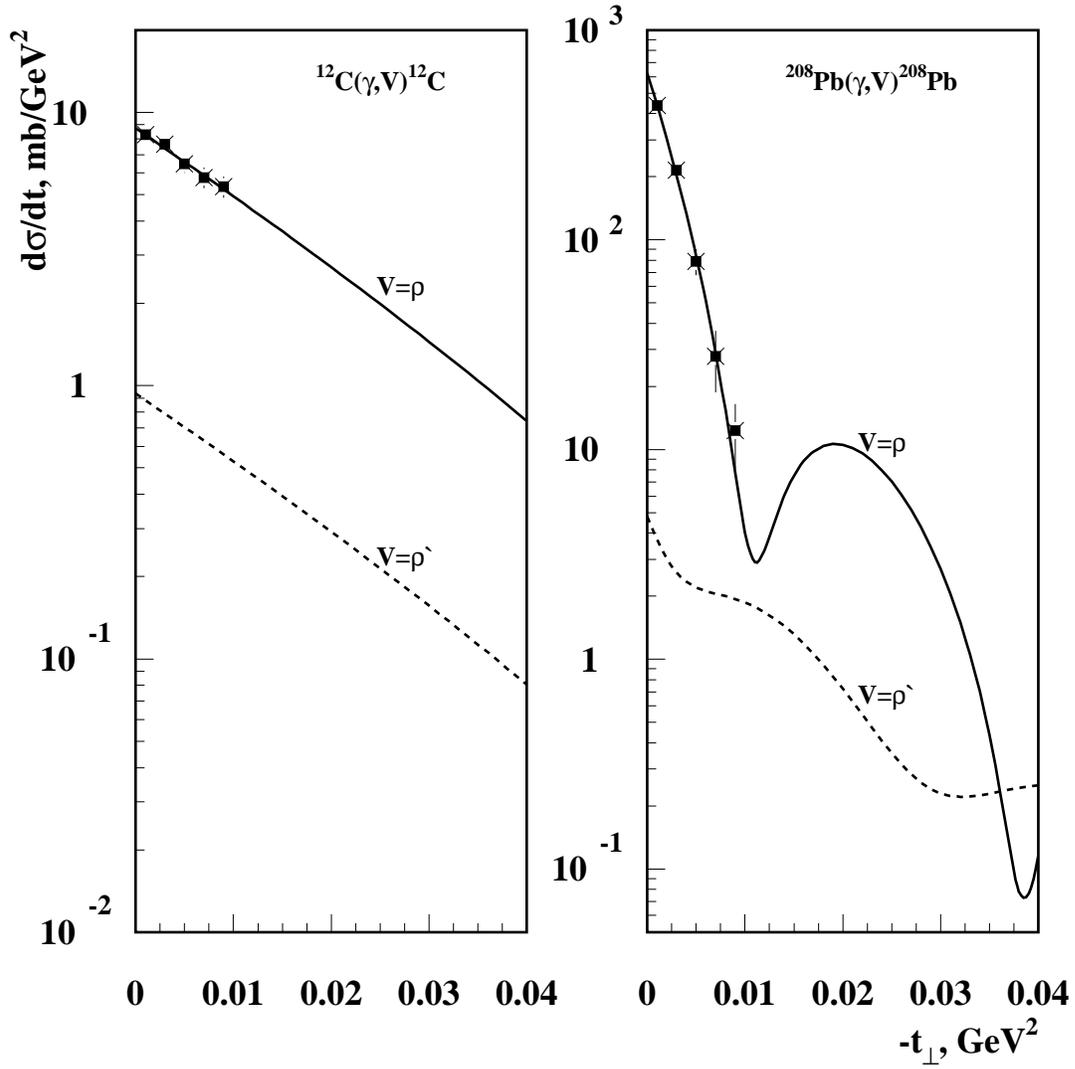}
\caption{The momentum transfer distributions of 
the $\rho$ and ${\rho^\prime}$
photoproduction  at 6.6 GeV calculated in the GVDM+Glauber model.}
\label{andis}
\end{figure}

\begin{figure}
\includegraphics[scale=0.8]{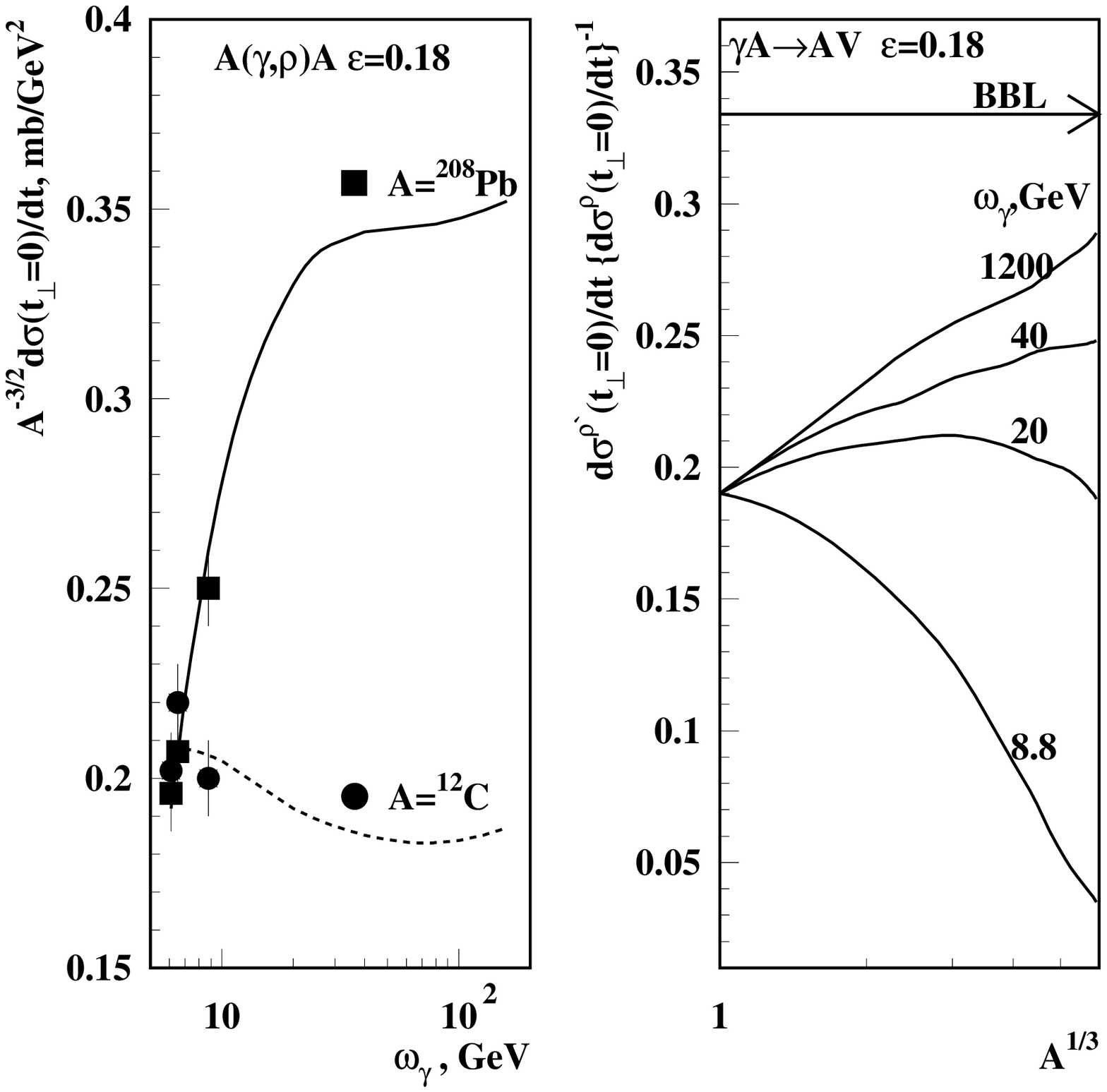}
\caption{The energy dependence of the $\rho$-photoproduction cross section
and the A-dependence of the $\rho^{\prime}/\rho$ photoproduction cross
sections calculated in the GVDM+Glauber model.}
\label{epend}
\end{figure}

We refined this model in \cite{FSZrho}. The dependence on the nuclear 
structure parameters was diminished by calculating the nuclear densities
in the Hartree-Fock-Skyrme  (HFS) approach which provided an excellent
description of the nuclear root mean square radii and the binding
energies of spherical nuclei for $A\ge 12$ and described well the
nuclear transparency in the high energy (p,2p) and (e,e'p) reactions, 
see \cite{eep} and refs. therein. Next, we  used in all our
calculations the parametrization of\cite{LD} for the  $\rho N$ amplitude 
which was obtained from the fit to the experimental data on photoproduction
off the proton target. The value of  $\epsilon $ was fixed at 0.18 to ensure 
the best fit of the measured differential cross section of
the $\rho$-meson photoproduction off lead at $\omega_{\gamma}=6.2$ GeV and
$p_t^2=0.001$ $GeV^2$  \footnote{The value $\epsilon=0.18$ leads to a 
suppression of the differential cross section of the $\rho$-photoproduction 
in $\gamma +p\to \rho +p$ by a factor of $(1-\epsilon/\sqrt{3})^2\approx 0.80$
practically coinciding with phenomenological  renormalization factor 
${ R=0.84}$ introduced in \cite{LD} to achieve the best fit of the elementary 
$\rho$-meson photoproduction forward cross section in the VDM which neglects 
mixing effects.}.
With all parameters fixed we calculated the  differential cross sections 
of $\rho$-production off nuclei and found a good agreement with all
available data \cite{MIT}, see Fig.\ref{andis} and 
a detailed comparison in \cite{FSZrho}.
In difference from \cite{Pautz:qm} we do not find it necessary
to increase the value of  $\epsilon$ by about $30\%$
when  the photon energy grows from 6 GeV to 8.8 GeV. As far as we know
previously this important check of the Glauber model predictions in the 
vector meson production off $A > 2$ nuclei(including the t-dependence of 
the cross section) has never  been performed in  such a  self-consistent way.
In view of a good agreement of the model with the data
on $\rho$-meson production in the low energy domain we  used this model
to consider the $\rho$ meson photoproduction at higher energies of
photons. The increase of the coherence length  with the photon energy
leads to a  qualitative difference in the energy dependence of
the coherent vector meson production off light
and heavy nuclei (Fig. \ref{epend})  and to a  change of the A-dependence
for the ratio of the forward $\rho^{\prime}$ and $\rho$-meson production
 cross sections between $\omega_{\gamma} \sim 10$ GeV 
and $\omega_{\gamma}\sim 40$ GeV (Fig.\ref{epend}).  
The observed pattern reflects the difference of  the 
coherence lengths of the $\rho$-meson and a heavier $\rho^\prime$-meson
which is important for the intermediate photon energies $\leq 30$ GeV.
The corrections due to nondiagonal transitions are relatively  small 
($\sim$ 15\%) for the case of  $\rho$ production off a nuclei. As a
result we find that the GVD cross section is close to the one calculated 
in the VD model for heavy nuclei as well. Situation is much more interesting 
for $\rho^{\prime}$ production.  The cross section of $\rho^{\prime}$ 
production off a nucleon is strongly suppressed as compared to the  case 
when the $\rho \leftrightarrow  \rho^{\prime}$  transitions are switched off.
The extra suppression factor is $\approx 0.5$.
In accordance with the general argument of Gribov the non-diagonal
transitions disappear in the limit of large A(black body limit)  due to 
the condition of orthogonality of hadronic wave functions \cite{Gribov}. 
Hence we expect that in the limit of $A\to \infty$ a more or less general 
relation 
\begin{equation}
{d\sigma (\gamma + A \to h_1 +A )/dt \over
d\sigma (\gamma + A \to h_2 +A )/dt}_{\left|A\to \infty\right.}
={\sigma(e^+e^-\to h_1)\over \sigma(e^+e^-\to h_2)}
\approx \left(f_2/f_1\right)^2
\label{bbl1}
\end{equation}
should be fulfiled for the productions of states $h_1, h_2$ of invariant 
masses $M_1^2,M_2^2$ at $k_t=0$.
Indeed we have found from calculations that in the case of the coherent 
photoproduction off lead the nondiagonal transitions becomes strongly
suppressed with increase of the photon energy. As  a result  the 
$\rho^{\prime}/\rho$ ratio increases, exceeds the ratio of the 
$\gamma p\to Vp$ forward cross sections calculated with accounting for 
$\rho-\rho^{\prime}$ transitions already at $\omega_{\gamma}\geq 50$ GeV 
and becomes close to  the value of $f^2_{\rho}/f^2_{\rho^{\prime}}$ which 
can be considered as the limit when one can treat the interaction with
the heavy nucleus as a  black one. 
It is worth noting here that presence of nondiagonal transitions
which in terms of the formalism of the scattering eigen states
\cite{GW} corresponds to the fluctuations of the values of the
interaction cross sections can be  even for real photon.
The GVDM model discussed in the paper leads to small ($\propto 10\%$) color 
transparency effects at intermediate energies for the cross section of
semiinclusive photoproduction processes. Really this model corresponds
to the propagation of states with cross sections:
$\approx \sigma(VN)(1 \pm \epsilon)$ . In the case of electroproduction 
$\epsilon$ should be significantly larger: 
$$\epsilon\approx{f_{\rho}\over f_{\rho'}}=
\sqrt{{\Gamma(\rho\rightarrow e^+ e^-)\over m_{\rho}}/
{\Gamma(\rho'\rightarrow e^+ e^-)\over m_{\rho'}}}.$$
This equation follows from Eq.(\ref{GVDM}) where left hand side
is put to 0 because cross section of elastic vector meson electroproduction 
rapidly decreases with  $Q^2$. The presence of the CT phenomenon within 
the GVDM leads to a substantial modification of the  pattern of the approach 
to BBL. With increase of $Q^2$ the nondiagonal transitions become more 
important leading to an enhancement of the effects discussed above. 
In particular found in the paper fluctuations of strengths of
interaction would lead at large $Q^2$ to the color transparency
phenomenon. Presence in GVDM of masses of $\rho,\rho'$ makes it
unreasonable and impossible to describe all fluctuations of strengths
of interaction in terms of one coherent length.
\section{LARGE MASS DIFFRACTION IN THE BLACK BODY LIMIT}
One of the striking features of the BBL is the suppression of nondiagonal
transitions in the photon interaction with heavy nuclei \cite {Gribov}.
Indeed in the BBL  the dominant contribution to the coherent  diffraction
originates from ``a shadow'' of the fully absorptive interactions
at impact parameters $b\leq R_A$ and hence the orthogonality condition 
is applicable. Using this argument it is easy to derive 
the BBL expression for the differential cross section of the 
production of the invariant mass $M^2$  for scattering of (virtual)
photons of transverse and longitudinal polarization:
\begin{equation}
{{d\sigma_{(\gamma +A\to ``M''+A)}}
\over dt dM^2} ={\alpha_{em} \over 3 \pi}{(2\pi R_A^2)^2\over 16\pi}
{\rho(M^2)\over M^2} {4\left|J_1(\sqrt{-t}R_A)\right|^2
\over -t R_A^2}.
\label{ccsb}
\end{equation}
Here $\rho(M^2) = 
\sigma(e^+ e^-\rightarrow hadrons)/ \sigma(e^+ e^-\rightarrow \mu^+ \mu^- )$.
Hence by comparing the extracted cross section of the diffractive
production of states with certain masses with the black body limit( 
Eq.(\ref{ccsb})) one would be able to determine up to what masses in
the photon wave function interaction remains black.
Actually similar equation is valid in the BBL for the production of specific
hadronic or quark-gluonic final states ($q\bar q, q\bar q g,$ etc)
in the case of the coherent nuclear recoil. Thus any component of the light 
cone photon wf whose interaction cross section leads to BBL is measurable by 
selecting certain final states in the coherent processes. 
Onset of BBL limit for hard processes should reveal itself also in the
faster increase with energy of cross sections of photoproduction of
excited states comparing to that for ground state meson.
It would be especially advantageous for these studies
to use a set of nuclei - one medium range like $Ca$ and
another heavy one - one could remove the edge effects and use the
length of about 10 fm of nuclear matter.

One especially interesting channel is 
diffractive dijet production by real photons.
For the $\gamma A$ energies which will be available at EIC or at LHC in UPC 
one may expect  that the BBL  in the scattering off heavy
nuclei would be a good approximation for the masses $M$ in the photon
wave function  up to  few GeV. This is 
the domain which is described by perturbative QCD for
${\it x}\sim 10^{-3}$ for the  proton targets and  larger $x$ 
for scattering off nuclei. The condition of large longitudinal 
distances - small longitudinal momentum 
transfer will be applicable in this case up to quite large values of the
produced diffractive mass.
In the BBL the dominant channel of diffraction for large masses is
production of two jets with the total cross section given by
Eq.(\ref{ccsb}) and with a characteristic angular distribution
$(1+ \cos^2 \theta)$, where $\theta$ is the c.m. angle \cite{BBL}. On
the contrary in the perturbative QCD limit the
diffractive dijet production except charmed jet production
is strongly suppressed \cite{brodsky,diehl}. The suppression
is due to the structure of coupling of the
real photon to two gluons when calculated in the lowest order in 
$\alpha_s$. As a result  in the real photon case hard diffraction
involving light quarks is connected to production of $q\bar q g$ and 
higher states. Distribution of diffractively produced jets over invariant mass 
provides an important  test of 
the onset of BBL limit. Really in the DGLAP/CT regime 
differential cross section of forward diffractive dijet production
should be $\propto 1/M^8$ and be dominated by charm jet production. This 
behaviour is strikingly different from BBL limit expressions of \cite{BBL}.
Thus the dijet photoproduction should be very sensitive to the
onset of the BBL regime. 
We want to draw attention that $q \bar q$ component
of the  photon light-cone wf can be measured in three
 independent diffractive phenomena:
in the BBL off the proton, in BBL off a heavy nuclei, in
CT regime where the wf can be measured as a 
function of the  interquark distance\cite{FMS}.
A competing process for photoproduction of dijets off heavy nuclei is
production of dijets in $\gamma -\gamma$ collisions where the second photon
is provided by the Coulomb field of the nucleus. 
The dijets produced in this process have positive
C-parity and hence this amplitude does not  interfere
with the amplitude of the dijet production in
the $\gamma \Pomeron$ interaction which have negative C-parity.
Our estimates indicate that this process will constitute a very small
background over the wide range of energies \cite{FSZrho}.

\begin{figure}
\includegraphics[scale=0.8]{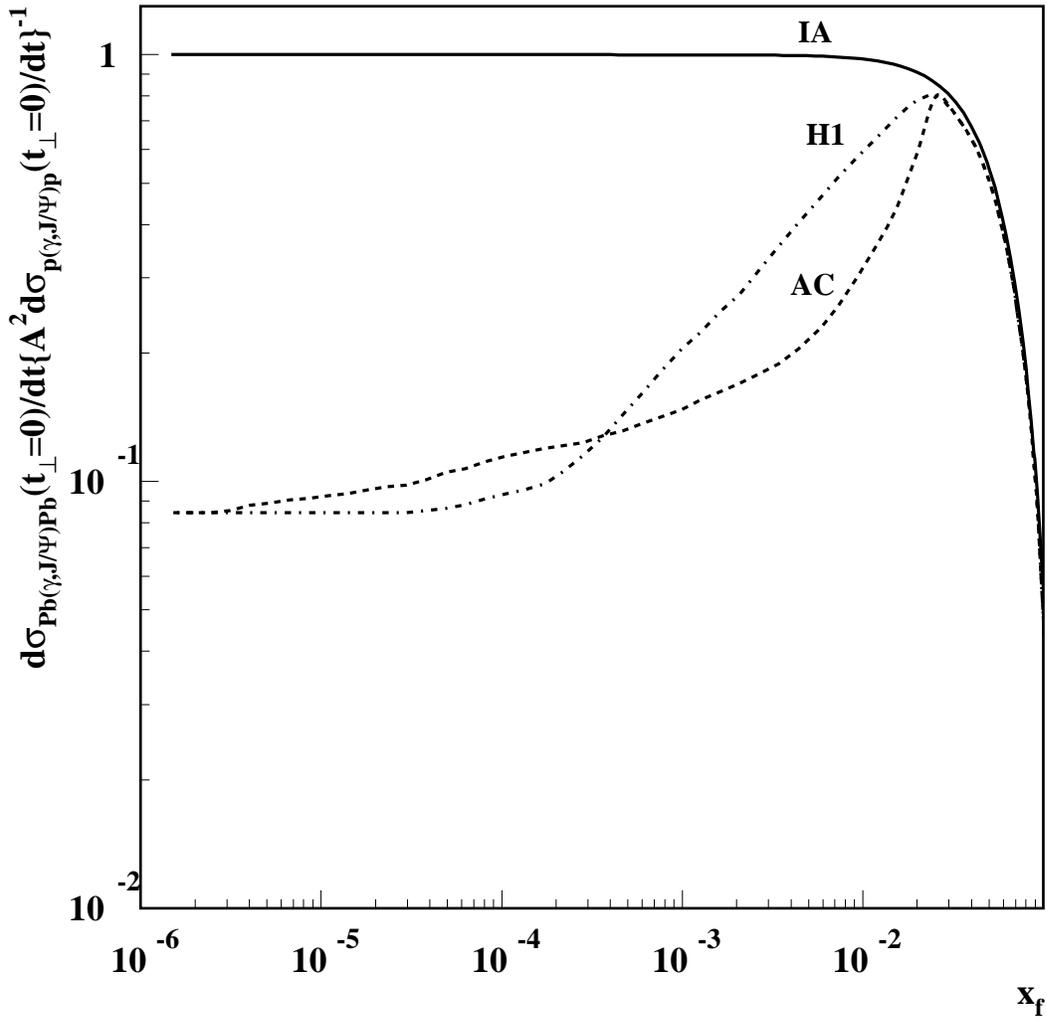}
\caption{The ${\it x}$-dependence of the ratio 
of $J/\psi$ production off Pb and a
nucleon. Solid line - calculation in Impulse Approximation(IA),
dashed line - calculation with the 
diffractive PDF of Alvero et.al.\cite{Alvero}, 
dot-dashed line -
H1 parametrization\cite{H1} of the diffractive PDF.}
\label{ferarj}
\end{figure}

\section{EXCLUSIVE MESON PRODUCTION AT LARGE $Q^2$.}
The QCD factorization theorem 
for exclusive meson production\cite{FMS,BFGMS} allows 
to express the amplitude   of the production of a vector meson
by a longitudinally polarized photon $\gamma_L + T \to V + T$
through the convolution of the wave function of the meson at 
the zero transverse separation, hard interaction block and the skewed 
parton density.
For practical purposes a crucial question is at what $Q^2$ 
squizing becomes effective. Probably the most sensitive indicator 
is the $t$-dependence of the meson production. The current HERA data
are consistent with the prediction of \cite{BFGMS,FKS}
that the slopes of the $\rho$ and $J/\psi$ production amplitudes 
should converge to the same value.
This indicates that at small ${\it x}$
configurations much smaller than average configurations
dominate for the $J/\psi$ production for all $Q^2$ and for $\rho$
for $Q^2\ge 5$ GeV$^2$.
In this kinematics the factorization theorem predicts
that the cross sections of coherent producton of vector mesons off
nuclei are proportional
(up to corrections due to skewedness effects
which for large $Q^2$ and small ${\it x}$ 
are calculable in terms of QCD evolution equation for skewed parton
distributions) to the square of the gluon parton density $G_A(x,Q^2)$ at 
small ${\it x}$ which is screened in nuclei as compared to  the nucleon.
Hence, one expects the regime of color transparency
for ${\it x}\ge 0.03$ where the gluon shadowing is very small/absent. At the 
same time at smaller ${\it x}$ one expects a gradual disappearance of color 
transparency \cite{FMS,BFGMS} - the  onset of the
perturbative color opacity.
 As an illustration we present in Fig.\ref{ferarj} the calculated 
${\it x}$-dependence of the transparency for $J/\psi$
production based on the leading twist gluon shadowing found in
\cite{FS99} where results for the
 $\Upsilon$ production are also presented.

\section{CONCLUSIONS}
We demonstrated that coherent diffraction off nuclei provides an
effective method of probing onset of BBL regime in hard processes at 
small ${\it x}$. We predict a significant  increase of the ratio of the 
yields of $\rho,\rho^{\prime}$ mesons in coherent processes off heavy 
nuclei due to the blackening of the soft QCD interactions in which 
fluctuations of the interaction strength are present. An account of 
nondiagonal transitions leads to a prediction of a significant enhancement 
of production of heavier diffractive states especially production of high 
$p_t$ dijets. Study of these channels may allow to get an important 
information on the onset of the black body limit in the diffraction of real 
photons. We argued that  the  fluctuations of strengths of 
interactions has been observed at intermediate energies in the diffractive 
photoproduction of vector mesons.

We thank J.Bjorken, S.Brodsky, A.Mueller, G.Shaw for  useful
discussions and GIF, CRDF and DOE for support.

\end{document}